\documentclass[aps,prc,twocolumn,floatfix,showpacs,a4paper,
nofootinbib,amsmath,amssymb]{revtex4-1}
\usepackage{graphicx}
\usepackage{dcolumn}
\usepackage{bm}
\usepackage{color}

\newcommand{\be}{\begin{equation}}
\newcommand{\ee}{\end{equation}}
\newcommand{\ba}{\begin{eqnarray}}
\newcommand{\ea}{\end{eqnarray}}
\newcommand{\bd}{\begin{displaymath}}
\newcommand{\ed}{\end{displaymath}}
\newcommand{\bea}{\begin{eqnarray}}
\newcommand{\eea}{\end{eqnarray}}

\begin{document}

\title{Nanoplasmonic Laser Fusion\\
Response to F\"oldes and Pokol\ - \ Letter to the Editor}

\author{
L\'aszl\'o P. Csernai\! $^{1}$,
Norbert Kro\'o\! $^{2,3}$
Istv\'an Papp\! $^2$,
Daniel D. Strottmam\! $^4$\\[1em]
$^1$ Dept. of Physics and Technology, Univ. of Bergen, Norway\\
$^2$ Wigner Research Centre for Physics, Budapest, Hungary\\
$^3$ Hungarian Academy of Sciences, Budapest, Hungary\\
$^4$ Los Alamos National Laboratory, Los Alamos, NM, USA\\
.\\
}

\date{ \today } 

\begin{abstract}
F\"oldes and Pokol
 in their letter "Inertial fusion without compression
does not work either with or without nanoplasmonics" criticized our works
\cite{CS2015,CKP2018}.
Here we refute their argumentation.
Our proposed improvement is the combination of two
basic research discoveries: (i) The possibility of detonations on space-time
hyper-surfaces with time-like normal (i.e. simultaneous detonation in a whole 
volume) and (ii) to increase the ignition volume to the whole target, 
by regulating the laser light absorption using nanoantennas.
These principles can be realized in an in-line, one dimensional configuration,
in the simplest way with two opposing laser beams as in particle colliders
\cite{CsEA2020,Bonasera2019}.
\end{abstract}

\maketitle

First of all the title of the letter of F\"oldes and Pokol is 
misleading. Already in the abstract of 
\cite{CKP2018} it is stated that we need to achieve limited 
compression to avoid the Rayleigh-Taylor instabilities. Furthermore
it is written that the aim is to achieve fusion "without too much 
pre-compression". It is false from F\"oldes and Pokol to state 
in their conclusion that we "... misjudged the possibility of 
the application of the uncompressed inertial fusion scheme.".
We have never had such a statement.

In 
\cite{CKP2018}
the LLNL NIF compression was discussed in some detail just as
it is done by F\"oldes and Pokol 
\cite{FP2020}.
The difficulties of this NIF setup by an ablator are elaborated in both of 
these publications.

F\"oldes and Pokol does not mention our work, where the improvements
we propose are elaborated in greater detail
\cite{CsEA2020}, and the experimental work showing that with the
simplified linear irradiation from both sides can achieve 
the same compression as reached at LLNL NIF
\cite{Bonasera2019}. These works provide a theoretical and experimental
response to the comments of F\"oldes and Pokol.

In recent years the Laser Wake Field Acceleration (LWFA) became a
well known concept with useful applications. An intensive laser pulse
impinging on a target creates a high density plasma of ~ 4 x 
10$^{19}$ / cm$^3$, and a wake field wave follows the pulse. 
This, non-linear wave in dense plasma is formed of the 
EM-field, electrons and ions. A typical laser pulse of 20 mJ intensity, 
7 fs length and $\lambda$ wave length can create a Laser Wake Field
(LWF) dense plasma wave of about 10 $\lambda$ wavelength. Electrons
do not prevent the development of this Laser Wake Field (LWF) waves.
In the configuration presented in 
\cite{CsEA2020,Bonasera2019}
such LWF waves develop and move opposite to each other like in
accelerators in collider configuration.
This is suggested in ref. 
\cite{CsEA2020}
for laser driven fusion. Here two known effects were combined. First, 
the possibility of detonations on space-time
hyper-surfaces with time-like normal, so called time-like detonations
\cite{Cs1987,CS2015} 
which were found theoretically and experimentally in high energy heavy
ion collision in the couple of last decades. This simultaneous volume
ignition eliminates the possibility of Rayleigh-Taylor instabilities,
which is a serious obstacle in achieving laboratory scale nuclear fusion.
The second effect we use is to achieve simultaneous detonation in the whole 
volume of the target, by regulating the laser light absorption with 
nano-shells or nano-rods.
This second effect is already experimentally proven in validation 
experiments, at Wigner RCP, Budapest, the institute of F\"oldes
\cite{KR2016}.

F\"oldes and Pokol under points 1., 2., 3., describe the thermonuclear 
setup of ICF where a dense hot-spot is created in $\approx$ 10-20 ns,
in the middle of the
target and then the flame front "slowly" propagates with 
$\alpha$-heating to the outside radius, $R$, of the target.
As we emphasize in all previous publications we do not follow this
dynamics, instead we aim to reach the condition for ignition
in $\approx$ 10-100 fs, in
most of the target volume, i.e. on a time-like hypersurface.
Then the flame front does not have to propagate anywhere, only
the local burning rate matters! In the short ignition time the ions
cannot expand.

This was also emphasized (even in the title) in the above mentioned,
published experiment using colliding laser beams
\cite{Bonasera2019}
achieved high target density for nuclear fusion.
On the other hand they did not utilize LWF waves and 
nanoplasmonics and did not
attempt to have simultaneous transition in the target. 
Thus, our present proposal differers from this experiment in the
two novel aspects.

In the publications commented on by F\"oldes and Polkol, we focused
on our two fundamental and novel ideas and did not discuss the
aspects, which should not be modified. 
In Laser Wake Field Collider, the target has two sides, which are
initially accelerated towards each other in a pre-acceleration and 
pre-compression process, on a nanosecond time scale,
just like at LLNL NIF and in 
\cite{Bonasera2019}. The LWFA may reach the GeV/nucleon energy
range.

In ref.
\cite{FLZhang2011}
it was shown that
intense laser pulse irradiating a combination target
can accelerate Carbon ions to the TeV level by the 
laser plasma wakefield.

If we consider LWFC with a double layer 
pre-compressed to ion density,  $n_{pc}$,
and pre-accelerated to several GeV/nucleon energy, (i.e. to a velocity
near to the speed of light, $v_{LWFC} \approx\ c$),
the two LWF waves can inter penetrate and lead to an ignition
reaction rate of
$$
     2 \ c\ \gamma^2\ n_{pc}^2\ \sigma \, ,
$$
due to the Lorentz contraction of the two ion bunches to $\gamma\ n_{pc}$,
and where $\sigma$ is the ion-ion cross section. 
This  may well exceed the thermal rate of $n \langle v_{th} \sigma \rangle $.
If we accelerate the ions to 5 GeV/nucleon, then $\gamma \approx 6$,
and if we achieve a pre-compression of factor 8 (considerably less
than at NIF, where the 3-Dim compression reaches 800 g/cm$^3$), then
our burning rate is 
$
2 \cdot \gamma^2\ \frac{\, 8^2 n^2}{800\,n}\ (c/v_{th}) \approx 270
$
times bigger than at NIF. Here we assumed that the average thermal 
collision speed is $v_{th} \approx c/2$.

So, the non-thermal non-equilibrium Laser Wake Field Collider mechanism
may well exceed the thermal ignition rate by the adiabatic compression
and heating at NIF. Especially if this ignition at NIF takes place for a
central hot-spot only and then the flame should propagate over the
rest of the target. The extreme energy need mentioned at the end of 
point 3., arises from the assumption of the thermal, setup, that we do not
follow.

According to our preliminary calculations (yet unpublished)
we can see that at around 125 - 150 fs after the target and
projectile touch, the two LWF waves constructively
interact and the EM field strength is maximal. This moment of time would 
be adequate for a short, intensive ignition pulse.

As we increase the target density 
the EM field penetration into the target is reduced, as the 
energy and momentum are passed to the target and projectile 
ions. The longitudinal momentum is transported
to the kinetic motion of the target particles, which then
contribute to the pinch effect reducing the EM strength
and target beam directed momentum. 

Under points 4., and 5., the target opacity and absorptivity are discussed.
Here we want to emphasize that our nanoantennas are embedded within
the target, so all energy reaching these remain in the target, 
so it is also absorbed. We have
no extreme losses that occur due to evaporating the external hohlraum.
The conversion of the absorbed energy to higher frequencies happens
within the nanoantennas or can lead to direct nuclear burning.

Under point 6. F\"oldes and Pokol assumes that the electrons will remain 
at the surface of the target. This fully contradicts with the structure 
of the LWF waves where the electrons are well separated from the ions.
The electrons within the target are concentrated in the
plasmonic waves on the surface of the nanoantennas. There is no significant
electron reflection observed in LWFA, and even if, these electrons would just 
again remain in the target fuel in the LWFC configuration. 

The conclusion of  F\"oldes and Pokol  that "Therefore the nanostructures 
inside the target are of no use at all" is already proven to be false,
as shown in the lower energy validation experiments at Wigner RCP, with
laser beams of up to 25 mJ pulse energy, and polymer targets, which exhibit
phase or structural transition at given threshold energy. This is
also discussed already in ref.
\cite{CsEA2020}
before the experiments were performed.

Summarizing the presented results both theoretically, and based on recent 
experiments, (both in fusion related and high energy heavy ion fields), we
can state that the conclusions of F\"oldes and Pokol are unsubstantiated. 


\end{document}